\documentclass[11pt,oneside]{article}
\usepackage{latexsym}
\usepackage{graphics}
\usepackage{amssymb}
\newcommand{\text}[1]{\mbox{#1}}
\newcommand{\binom}[2]{{#1 \choose #2}}
\newcommand{\QATOP}[2]{{#1 \atop #2}}

\begin{document}

\author{Hartmut Wachter\thanks{%
e-mail:Hartmut.Wachter@physik.uni-muenchen.de} \\
%EndAName
Sektion Physik, Ludwig-Maximilians-Universit\"{a}t,\\
Theresienstr. 37, D-80333 M\"{u}nchen, Germany}
\title{q-Exponentials on quantum spaces}
\date{}
\maketitle

\begin{abstract}
We present explicit formulae for q-exponentials on quantum spaces which
could be of particular importance in physics, i.e. the q-deformed Minkowski
space and the q-deformed Euclidean space with three or four dimensions.
Furthermore, these formulae can be viewed as 2-, 3- or 4-dimensional
analogues of the well-known q-exponential function.
\end{abstract}

%\documentclass[12pt,thmsa]{article}
%\usepackage{amsfonts}

%%%%%%%%%%%%%%%%%%%%%%%%%%%%%%%%%%%%%%%%%%%%%%%%%%%%%%%%%%%%%%%%%%%%%%%%%%%%%%%%%%%%%%%%%%%%%%%%%%%
%\usepackage{sw20lart}

%TCIDATA{TCIstyle=article/art4.lat,lart,article}

%TCIDATA{Created=Tue Nov 18 18:49:28 2003}
%TCIDATA{LastRevised=Thu May 13 22:31:19 2004}

%\input{tcilatex}
%\begin{document}

%\author{Hartmut Wachter\thanks{%
%e-mail:Hartmut.Wachter@physik.uni-muenchen.de} \\
%EndAName
%Sektion Physik, Ludwig-Maximilians-Universit\"{a}t,\\
%Theresienstr. 37, D-80333 M\"{u}nchen, Germany}
%\title{q-Exponentials on quantum spaces}
%\date{}
%\maketitle

%\begin{abstract}
%We present explicit formulae for q-exponentials on quantum spaces which
%could be of particular importance in physics, i.e. the q-deformed Minkowski
%space and the q-deformed Euclidean space with three or four dimensions.
%Furthermore, these formulae can be viewed as 2-, 3- or 4-dimensional
%analogues of the well-known q-exponential function.
%\end{abstract}

\section{Introduction}

In this work we would like to continue our programme for developing a
non-commutative analysis, which in the following is referred to as
q-analysis. One of our motivations for doing this is that a field theory
based on q-analysis should be well behaved in the UV-range
\cite{GKP96,Oec99,Blo03,ChDe94}. So far we have concerned ourselves
with explicit formulae for star-products \cite{WW01}, representations
of partial derivatives \cite{BW01} as well as q-integrals
\cite{Wac02}. All of these
mathematical objects have been considered for such non-commutative spaces
which could be of particular importance in physics, i.e. the q-deformed
Minkowski space and the q-deformed Euclidean space with three or four
dimensions. Our goal now is to take a step towards completing this programme
by providing also explicit formulae for q-exponentials.

Before doing this let us recall some basic aspects of our approach. As
already mentioned, q-analysis can be regarded as a non-commutative analysis
formulated within the framework of quantum spaces
\cite{Maj93-6,Maj94-10,Wess00}. These quantum spaces are defined as comodule
algebras of quantum groups and\ can therefore be interpreted as deformations
of ordinary coordinate algebras \cite{RTF90}. For our purposes it is at
first sufficient to consider a quantum space as an algebra $\mathcal{A}_{q}$
of formal power series in the non-commuting coordinates $X_{1},X_{2},\ldots
,X_{n}$

\begin{equation}
\mathcal{A}_{q}=\mathbb{C}\left[ \left[ X_{1},\ldots X_{n}\right] \right] /%
\mathcal{I},
\end{equation}
where $\mathcal{I}$ denotes the ideal generated by the relations of the
non-commuting coordinates. The algebra $\mathcal{A}_{q}$ satisfies the
Poincar\'{e}-Birkhoff-Witt property, i.e. the dimension of the subspace of
homogenous polynomials should be the same as for commuting coordinates. This
property is the deeper reason why the monomials of normal ordering $%
X_{1}X_{2}\ldots X_{n}$ constitute a basis of $\mathcal{A}_{q}$. In
particular, we can establish a vector space isomorphism between $\mathcal{A}%
_{q}$ and the commutative algebra $\mathcal{A}$ generated by ordinary
coordinates $x_{1},x_{2},\ldots ,x_{n}$: 
\begin{eqnarray}
\mathcal{W} &:&\mathcal{A}\longrightarrow \mathcal{A}_{q},  \label{AlgIso} \\
\mathcal{W}(x_{1}^{i_{1}}\ldots x_{n}^{i_{n}}) &=&X_{1}^{i_{1}}\ldots
X_{n}^{i_{n}}.  \nonumber
\end{eqnarray}
This vector space isomorphism can be extended to an algebra isomorphism
introducing a non-commutative product in $\mathcal{A}$, the so-called $\star 
$-product \cite{Moy49,MSSW00}. This product is defined by the
relation 
\begin{equation}
\mathcal{W}(f\star g)=\mathcal{W}(f)\cdot \mathcal{W}(g),
\end{equation}
where $f$ and $g$ are formal power series in $\mathcal{A}$. Additionally,
for each quantum space exists a symmetry algebra \cite{Dri85,Jim85}
and a covariant differential calculus \cite{WZ91} which can provide an
action upon the quantum spaces under consideration. By means of the relation 
\begin{equation}
\mathcal{W}(h\triangleright f):=h\triangleright \mathcal{W}(f)\text{,\quad }%
h\in \mathcal{H}\text{, }f\in \mathcal{A}\text{,}  \label{defrep}
\end{equation}
we are also able to introduce an action upon the corresponding commutative
algebra.

To gain further insight it is also useful to consider quantum spaces from a
point of view provided by category theory. A category for our purposes is
just a collection of objects $X,Y,Z,\ldots $ and a set Mor$(X,Y)$ of
morphisms between two objects $X,Y$ such that a composition of morphisms is
defined which has similar properties to the composition of maps. In
particular, we are interested in tensor categories. These are categories
that have a product, denoted $\otimes $ and called the tensor product, which
admits several 'natural' properties such as associativity and existence of a
unit object. For a more formal treatment we refer the interested reader to
the presentations in \cite{Maj91Kat}, \cite{Maj94Kat} or \cite{MaL74}.

Essentially for us is the fact that the representations (quantum spaces) of
a given quasitriangular Hopf algebra (quantum algebra) are the objects of a
tensor category, if the action of the Hopf algebra on the tensor product of
two quantum spaces is defined by\footnote{%
We write the coproduct in the so-called Sweedler notation, i.e. $\Delta
(h)=h_{(1)}\otimes h_{(2)}.$} 
\begin{equation}
h\triangleright (v\otimes w)=(h_{(1)}\triangleright v)\otimes
(h_{(2)}\triangleright w).
\end{equation}
In this category exist a number of morphisms of particular importance. First
of all, for any pair of objects $X,Y$ there is an isomorphism $\Psi
_{X,Y}:X\otimes Y\rightarrow Y\otimes X$ such that $(g\otimes f)\circ \Psi
_{X,Y}=\Psi _{X^{\prime },Y^{\prime }}\circ (f\otimes g)$ for arbitrary
morphisms $f\in $ Mor$(X,X^{\prime })$ and $g\in $ Mor$(Y,Y^{\prime })$ and
the hexagon axiom holds. This hexagon axiom is the validity of the two
conditions 
\begin{equation}
\Psi _{X,Y}\circ \Psi _{Y,Z}=\Psi _{X\otimes Y,Z},\quad \Psi _{X,Z}\circ
\Psi _{X,Y}=\Psi _{X,Y\otimes Z}.
\end{equation}
A tensor category having the above property is called a braided tensor
category. Furthermore, for any algebra $B$ in this category there are
morphisms $\Delta :B\rightarrow B$\underline{$\otimes $}$B,$ $S:B\rightarrow
B$ and $\varepsilon :B\rightarrow \mathbb{C}$ forming a braided Hopf algebra,
i.e. $\Delta ,$ $S$ and $\epsilon $ obey the usual axioms of a Hopf algebra,
but now as morphisms in the braided category. These considerations show that
under suitable assumptions our quantum spaces can be viewed as braided
Hopf algebras. For a deeper understanding of these ideas we also refer
the reader to the excellent presentations in \cite{ChDe96} and \cite{MajBuch}.

Now let us make contact with another very important ingredient of our
braided tensor category. For this purpose we suppose that our category is
equipped with dual objects $B^{*}$ for each algebra $B$ in the category.
This means that we have a dual pairing 
\begin{equation}
\left\langle \,,\,\right\rangle :B\otimes B^{*}\rightarrow K\quad \text{%
with\quad }\left\langle e_{a},f^{b}\right\rangle =\delta _{a}^{b},
\end{equation}
where $\{e_{a}\}$ is a basis in $B$ and $\{f_{a}\}$ a dual basis in $B^{*}.$
This allows us to introduce an exponential which from an abstract point of
view is nothing other than an object whose dualisation is the evaluation map 
\cite{Maj94-10}. Thus, the exponential is given by the map\footnote{%
In the quantum group case such an object is often referred to as the
canonical element.} 
\begin{equation}
\exp :K\rightarrow B^{*}\otimes B\quad \text{with\quad }\exp
=\sum_{a}f^{a}\otimes e_{a}  \label{ExpAll}
\end{equation}
and satisfies the following relations 
\begin{eqnarray}
(\bigtriangleup \otimes id)\exp &=&\sum_{j,k}f^{j}\otimes f^{k}\otimes
e_{k}e_{j}=\exp _{23}\exp _{13},  \label{KomGes} \\
(id\otimes \bigtriangleup )\exp &=&\sum_{j,k}f^{j}f^{k}\otimes e_{k}\otimes
e_{j}=\exp _{13}\exp _{12}.  \nonumber
\end{eqnarray}

To make this concrete, we recall that it was shown in \cite{Maj93-5} that
there is such a duality pairing of quantum space coordinates and the
corresponding partial derivatives. Explicitly, we have 
\begin{equation}
\left\langle \,,\,\right\rangle :\mathcal{M}_{\partial }\otimes \mathcal{M}%
_{x}\rightarrow K\quad \text{with\quad }\left\langle f(\partial
_{i}),g(x^{j})\right\rangle \equiv \varepsilon (f(\partial
_{i})\triangleright g(x^{j})).
\end{equation}

As it is our aim to derive explicit formulae for q-exponentials of
q-deformed Minkowski space and q-deformed Euclidean space with up to four
dimensions, our task is therefore to determine a basis of the coordinate
algebra $\mathcal{M}_{x}$ being dual to a given one of the derivative
algebra $\mathcal{M}_{\partial }$. Inserting the elements of these two basis
into formula (\ref{ExpAll}) will then provide us with explicit expressions
for the exponentials. It should be stressed that the existence of the
algebra isomorphism $\mathcal{W}$ defined in (\ref{AlgIso}) enables us to
carry out all the necessary calculations in the corresponding commutative
algebras. In doing so we are led to 2-, 3- and 4-dimensional analogues of
the well-known q-exponential function \cite{BL95,Koo96}.

As it was shown in \cite{BW01}, the partial derivatives can act on the
algebra of quantum space coordinates in four different ways, i.e. by
right-actions, left-actions and their conjugated counterparts. For this
reason there are four possibilities for defining a pairing between
coordinates and derivatives. More concretely, we can distinguish the
pairings 
\begin{eqnarray}
\langle f(\partial ),g(x)\rangle _{L,\hat{R}} &\equiv &\varepsilon
(f(\partial )\triangleright g(x)),  \label{AllPaar} \\
\langle f(\hat{\partial}),g(x)\rangle _{\hat{L},R} &\equiv &\varepsilon (f(%
\hat{\partial})\,\bar{\triangleright}\,g(x)),  \nonumber \\
\langle f(x),g(\partial )\rangle _{L,\hat{R}} &\equiv &\varepsilon (f(x)\,%
\bar{\triangleleft}\,g(\partial )),  \nonumber \\
\langle f(x),g(\hat{\partial})\rangle _{\hat{L},R} &\equiv &\varepsilon
(f(x)\triangleleft g(\hat{\partial})),  \nonumber
\end{eqnarray}
where $\hat{\partial}^{A}$ differs from $\partial ^{A}$ by a normalisation
factor, only \cite{BW01,LWW97,CSW91}. Clearly, each of these
pairings will lead to its own exponential. It should also be clear from the
above considerations that the different exponentials can be linked via the
same crossing symmetries which have already helped us in \cite{BW01} to
transform the underlying representations of partial derivatives into each
other.

There are two properties of the considered exponentials worth recording
here. First of all, the exponentials are normalized in such a way that 
\begin{eqnarray}
(\varepsilon \otimes id)\exp (x\mid \partial ) &=&1, \\
(id\otimes \varepsilon )\exp (x\mid \partial ) &=&1.  \nonumber
\end{eqnarray}
These equalities result from 
\begin{eqnarray}
(id\otimes \varepsilon )\exp  &=&\sum_{a}f^{a}\otimes \varepsilon (e_{a})
\label{NorEbWel} \\
&=&\sum_{a}f^{a}\otimes \langle e_{a},1\rangle   \nonumber \\
&=&1\otimes 1=1,  \nonumber \\[0.16in]
(\varepsilon \otimes id)\exp  &=&\sum_{a}\varepsilon (f^{a})\otimes e_{a} \\
&=&\sum_{a}\langle 1,f^{a}\rangle \otimes e_{a}  \nonumber \\
&=&1\otimes 1=1.  \nonumber
\end{eqnarray}
Second the exponentials obey the identities \cite{MajBuch} 
\begin{eqnarray}
\partial ^{i}\triangleright \exp (x_{\hat{R}}\mid \partial _{L}) &=&\exp (x_{%
\hat{R}}\mid \partial _{L})\star \partial ^{i},  \label{EigGl} \\
\hat{\partial}^{i}\,\bar{\triangleright}\,\exp (x_{R}\mid \hat{\partial}_{%
\hat{L}}) &=&\exp (x_{R}\mid \hat{\partial}_{\hat{L}})\star \hat{\partial}%
^{i},  \nonumber \\
\partial ^{i}\star \exp (\partial _{\hat{R}}\mid x_{L}) &=&\exp (\partial _{%
\hat{R}}\mid x_{L})\,\bar{\triangleleft}\,\partial ^{i},  \nonumber \\
\hat{\partial}^{i}\star \exp (\hat{\partial}_{R}\mid x_{\hat{L}}) &=&\exp (%
\hat{\partial}_{R}\mid x_{\hat{L}})\triangleleft \hat{\partial}^{i} 
\nonumber
\end{eqnarray}
which tell us that our exponentials can be regarded as q-analogues of
classical plane-waves. For a proof of these formulae one has to realize that
an algebra acts on its dual via 
\begin{equation}
e_{a}\triangleright f^{b}=\langle e_{a},f_{(1)}^{b}\rangle f_{(2)}^{b}.
\end{equation}
With this relation at hand one can proceed in the following fashion: 
\begin{eqnarray}
&&(e_{b}\otimes 1)\triangleright \exp  \\
&=&\sum_{a}e_{b}\triangleright f^{a}\otimes e_{a}  \nonumber \\
&=&\sum_{a}\langle e_{b},f_{(1)}^{a}\rangle f_{(2)}^{a}\otimes e_{a} 
\nonumber \\
&=&\sum_{a,c}\langle e_{b},f^{a}\rangle f^{c}\otimes e_{c}e_{a}  \nonumber \\
&=&\sum_{c}f^{c}\otimes e_{c}e_{b}  \nonumber \\
&=&\exp \cdot (1\otimes e_{b}).  \nonumber
\end{eqnarray}
Notice that the third equality uses the first relation in (\ref{KomGes}).
Through a slight modification to these arguments, one can proof the
corresponding identities concerning right actions.

\section{2-Dimensional q-deformed Euclidean space}

The q-deformed Euclidean space with two dimensions is generated by
coordinates $X^{i},$ $i=1,2,$ subject to the relation 
\begin{equation}
X^{1}X^{2}=qX^{2}X^{1}.
\end{equation}
As it is well-known, there are two covariant differential calculi on this
quantum space given by \cite{WZ91} 
\begin{eqnarray}
\partial ^{i}X^{j} &=&\varepsilon ^{ij}+q^{2}(\hat{R}^{-1})_{kl}^{ij}X^{k}%
\partial ^{l}, \\
\hat{\partial}^{i}X^{j} &=&\varepsilon ^{ij}+q^{-2}(\hat{R}%
)_{kl}^{ij}X^{k}\partial ^{l},\quad i,j=1,2,  \nonumber
\end{eqnarray}
where $\hat{R}$ and $\varepsilon ^{ij}$ denote respectively the R-matrix of
the quantum algebra $U_{q}(su_{2})$ and the corresponding quantum metric%
\footnote{%
Our notations, conventions and definitions are listed in the appendix.}.
Written out, we get 
\begin{eqnarray}
\partial _{X}^{1}X^{1} &=&qX^{1}\partial _{X}^{1},  \label{ExVerAbKoanf} \\
\partial _{X}^{1}X^{2} &=&-q^{-\frac{1}{2}}+q^{2}X^{2}\partial _{X}^{1}, 
\nonumber \\[0.16in]
\partial _{X}^{2}X^{1} &=&q^{\frac{1}{2}}+q^{2}X^{1}\partial
_{X}^{2}-q^{2}\lambda X^{2}\partial _{X}^{1},  \label{ExVerAblKoend} \\
\partial _{X}^{2}X^{2} &=&qX^{2}\partial _{X}^{2}  \nonumber
\end{eqnarray}
and 
\begin{eqnarray}
\hat{\partial}_{X}^{1}X^{1} &=&q^{-1}X^{1}\hat{\partial}_{X}^{1}, \\
\hat{\partial}_{X}^{1}X^{2} &=&q^{-\frac{1}{2}}+q^{-2}X^{2}\hat{\partial}%
_{X}^{1}+q^{-2}\lambda X^{1}\hat{\partial}_{X}^{2},  \nonumber \\[0.16in]
\hat{\partial}_{X}^{2}X^{1} &=&-q^{\frac{1}{2}}+q^{-2}X^{1}\hat{\partial}%
_{X}^{2}, \\
\hat{\partial}_{X}^{2}X^{2} &=&q^{-1}X^{2}\hat{\partial}_{X}^{2},  \nonumber
\end{eqnarray}
with $\lambda=q-q^{-1}$.
From these commutation relations one can calculate the action of the partial
derivatives on monomials of a given normal ordering. In this way we have
derived the expressions 
\begin{eqnarray}
\partial ^{1}\triangleright (X^{2})^{m_{2}}(X^{1})^{m_{1}} &=&-q^{-\frac{1}{2%
}}[[m_{2}]]_{q^{2}}(X^{2})^{m_{2}-1}(X^{1})^{m_{1}}, \label{actpartwo}\\
\partial ^{2}\triangleright (X^{2})^{m_{2}}(X^{1})^{m_{1}} &=&q^{\frac{1}{2}%
+m_{2}}[[m_{1}]]_{q^{2}}(X^{2})^{m_{2}}(X^{1})^{m_{1}-1}  \nonumber
\end{eqnarray}
and 
\begin{eqnarray}
\hat{\partial}^{1}\,\bar{\triangleright}\,(X^{1})^{m_{1}}(X^{2})^{m_{2}}
&=&q^{-\frac{1}{2}-m_{1}}[[m_{2}]]_{q^{-2}}(X^{1})^{m_{1}}(X^{2})^{m_{2}-1},
\label{actpartend}\\
\hat{\partial}^{2}\,\bar{\triangleright}\,(X^{1})^{m_{1}}(X^{2})^{m_{2}}
&=&-q^{\frac{1}{2}}[[m_{1}]]_{q^{-2}}(X^{1})^{m_{1}-1}(X^{2})^{m_{2}}, 
\nonumber
\end{eqnarray}
where the antisymmetric q-number is defined by \cite{KS97} 
\begin{equation}
\left[ \left[ c\right] \right] _{q^{a}}\equiv \frac{1-q^{ac}}{1-q^{a}}%
,\qquad a,c\in \mathbb{C}.
\end{equation}
Iteration of (\ref{actpartwo}) and (\ref{actpartend}) then leads to
the formulae 
%\begin{eqnarray}
%&&(\partial ^{2})^{n_{2}}(\partial ^{1})^{n_{1}}\triangleright
%(X^{2})^{m_{2}}(X^{1})^{m_{1}} \\
%&=&(-q^{-\frac{1}{2}})^{n_{1}}q^{\frac{1}{2}%
%n_{2}}q^{n_{2}(m_{2}-n_{1})}[[n_{1}]]_{q^{2}}![[n_{2}]]_{q^{2}}!  \nonumber
%\\
%&&\cdot \QATOPD {m_{2}}{n_{1}}_{q^{2}}\QATOPD {m_{1}}{n_{2}}%
%_{q^{2}}(X^{2})^{m_{2}-n_{1}}(X^{1})^{m_{1}-n_{2}},  \nonumber \\[0.16in]
%&&(\hat{\partial}^{1})^{n_{1}}(\hat{\partial}^{2})^{n_{2}}\,\bar{%
%\triangleright}\,(X^{1})^{m_{1}}(X^{2})^{m_{2}} \\
%&=&(-q^{\frac{1}{2}})^{n_{2}}q^{-\frac{1}{2}%
%n_{1}}q^{-n_{1}(m_{1}-n_{2})}[[n_{1}]]_{q^{-2}}![[n_{2}]]_{q^{-2}}! 
%\nonumber \\
%&&\cdot \QATOPD {m_{1}}{n_{2}}_{q^{-2}}\QATOPD {m_{2}}{n_{1}}%
%_{q^{-2}}(X^{1})^{m_{1}-n_{2}}(X^{2})^{m_{2}-n_{1}},  \nonumber
%\end{eqnarray}
%which, in turn, give
for the dual pairing\footnote{%
Notice, that $\varepsilon (f(\underline{x}))=f(\underline{x}=0).$} 
\begin{eqnarray}
&&\Big\langle (\partial ^{2})^{n_{2}}(\partial
^{1})^{n_{1}},(X^{2})^{m_{2}}(X^{1})^{m_{1}}\Big\rangle _{L,\hat{R}} \\
&=&\varepsilon \Big ((\partial ^{2})^{n_{2}}(\partial
^{1})^{n_{1}}\triangleright (X^{2})^{m_{2}}(X^{1})^{m_{1}}\Big)  \nonumber \\
&=&\delta _{m_{1},n_{2}}\delta _{m_{2},n_{1}}(-q^{-\frac{1}{2}})^{n_{1}}q^{%
\frac{1}{2}n_{2}}[[n_{1}]]_{q^{2}}![[n_{2}]]_{q^{2}}!,  \nonumber \\[0.16in]
&&\Big\langle (\hat{\partial}^{1})^{n_{1}}(\hat{\partial}%
^{2})^{n_{2}},(X^{1})^{m_{1}}(X^{2})^{m_{2}}\Big\rangle _{\hat{L},R} \\
&=&\varepsilon \Big ((\hat{\partial}^{1})^{n_{1}}(\hat{\partial}%
^{2})^{n_{2}}\,\bar{\triangleright}\,(X^{1})^{m_{1}}(X^{2})^{m_{2}}\Big) 
\nonumber \\
&=&\delta _{m_{1},n_{2}}\delta _{m_{2},n_{1}}(-q^{\frac{1}{2}})^{n_{2}}(q^{-%
\frac{1}{2}})^{n_{1}}[[n_{1}]]_{q^{-2}}![[n_{2}]]_{q^{-2}}!,  \nonumber
\end{eqnarray}
where the q-factorials are given by
\begin{equation}
\left[ \left[ m\right] \right] _{q^{a}}!\equiv \left[ \left[ 1\right]
\right] _{q^{a}}\left[ \left[ 2\right] \right] _{q^{a}}\ldots \left[ \left[
m\right] \right] _{q^{a}},\qquad \left[ \left[ 0\right] \right]
_{q^{a}}!\equiv 1.
\end{equation}
The above results can be simplified further by introducing partial
derivatives with lower indices defined by 
\begin{equation}
\partial _{i}=\varepsilon _{ij}\partial ^{j},\quad \hat{\partial}%
_{i}=\varepsilon _{ij}\hat{\partial}^{j}.
\end{equation}
Thus, we end up with 
\begin{eqnarray}
&&\Big\langle (\partial _{1})^{n_{1}}(\partial
_{2})^{n_{2}},(X^{2})^{m_{2}}(X^{1})^{m_{1}}\Big\rangle _{L,\hat{R}} \\
&=&\delta _{m_{1},n_{1}}\delta
_{m_{2},n_{2}}[[n_{1}]]_{q^{2}}![[n_{2}]]_{q^{2}}!,  \nonumber \\[0.16in]
&&\Big\langle (\hat{\partial}_{2})^{n_{2}}(\hat{\partial}%
_{1})^{n_{1}},(X^{1})^{m_{1}}(X^{2})^{m_{2}}\Big\rangle _{\hat{L},R} \\
&=&\delta _{m_{1},n_{1}}\delta
_{m_{2},n_{2}}[[n_{1}]]_{q^{-2}}![[n_{2}]]_{q^{-2}}!.  \nonumber
\end{eqnarray}
These expressions enable us to identify the two sets of basis elements being
dual to each other. With this knowledge we are now in a position to apply
formula (6) giving us 
\begin{eqnarray}
\widetilde{\exp }(x_{\hat{R}}\mid \partial _{L})
&=&\sum_{n_{1},n_{2}=0}^{\infty }\frac{(X^{2})^{n_{2}}(X^{1})^{n_{1}}\otimes
(\partial _{1})^{n_{1}}(\partial _{2})^{n_{2}}}{%
[[n_{1}]]_{q^{2}}![[n_{2}]]_{q^{2}}!},  \label{exp2Dimanf} \\
\exp (x_{R}\mid \hat{\partial}_{\hat{L}}) &=&\sum_{n_{1},n_{2}=0}^{\infty }%
\frac{(X^{1})^{n_{1}}(X^{2})^{n_{2}}\otimes (\hat{\partial}_{2})^{n_{2}}(%
\hat{\partial}_{1})^{n_{1}}}{[[n_{1}]]_{q^{-2}}![[n_{2}]]_{q^{-2}}!},
\label{exp2Dimend}
\end{eqnarray}
where the tilde in the first formula shall remind us of the fact that this
exponential compared to the second one refers to a different choice for the
normal ordering of the coordinates and derivatives.

The remainder of this section is devoted to the calculation of exponentials
corresponding to right representations of partial derivatives. For this
purpose it is useful to realize that left representations are transformed to
right ones by conjugation, i.e. 
\begin{eqnarray}
\overline{\partial \triangleright f} &=&\bar{f}\,\bar{\triangleleft}\,\bar{%
\partial}, \\
\overline{\partial \,\bar{\triangleright}\,f} &=&\bar{f}\triangleleft \bar{%
\partial}.  \nonumber
\end{eqnarray}
In view of this relationship and the conjugation properties\footnote{%
We are here following the approach
of \cite{Maj95star} which implies that partial
derivatives and coordinates obey the same conjugation properties.} 
\begin{equation}
\overline{h^{i}}=\varepsilon _{ij}h^{j},
\end{equation}
where $h^{i}$ stands for $X^{i}$ or $\partial ^{i}$, the right actions of
partial derivatives on normally ordered monomials take on the form 
\begin{eqnarray}
(X^{2})^{m_{2}}(X^{1})^{m_{1}}\,\bar{\triangleleft}\,\partial ^{1} &=&-q^{%
\frac{1}{2}+m_{1}}[[m_{2}]]_{q^{2}}(X^{2})^{m_{2}-1}(X^{1})^{m_{1}}, \\
(X^{2})^{m_{2}}(X^{1})^{m_{1}}\,\bar{\triangleleft}\,\partial ^{2} &=&q^{-%
\frac{1}{2}}[[m_{1}]]_{q^{2}}(X^{2})^{m_{2}}(X^{1})^{m_{1}-1}  \nonumber
\end{eqnarray}
and 
\begin{eqnarray}
(X^{1})^{m_{1}}(X^{2})^{m_{2}}\triangleleft \hat{\partial}^{1} &=&q^{\frac{1%
}{2}}[[m_{2}]]_{q^{-2}}(X^{1})^{m_{1}}(X^{2})^{m_{2}-1}, \\
(X^{1})^{m_{1}}(X^{2})^{m_{2}}\triangleleft \hat{\partial}^{2} &=&-q^{-\frac{%
1}{2}-m_{2}}[[m_{1}]]_{q^{-2}}(X^{1})^{m_{1}-1}(X^{2})^{m_{2}}.  \nonumber
\end{eqnarray}
With the very same reasonings already applied to left representations we can
show that 
\begin{eqnarray}
&&\Big\langle (X_{1})^{m_{1}}(X_{2})^{m_{2}},(\partial
^{2})^{n_{2}}(\partial ^{1})^{n_{1}}\Big\rangle _{L,\hat{R}} \\
&=&\varepsilon ((X_{1})^{m_{1}}(X_{2})^{m_{2}}\,\bar{\triangleleft}%
\,(\partial ^{2})^{n_{2}}(\partial ^{1})^{n_{1}})  \nonumber \\
&=&\delta _{m_{1},n_{1}}\delta
_{m_{2},n_{2}}[[n_{1}]]_{q^{2}}![[n_{2}]]_{q^{2}}!,  \nonumber \\[0.16in]
&&\Big\langle (X_{2})^{m_{2}}(X_{1})^{m_{1}},(\hat{\partial}^{1})^{n_{1}}(%
\hat{\partial}^{2})^{n_{2}}\Big\rangle _{\hat{L},R} \\
&=&\varepsilon ((X_{2})^{m_{2}}(X_{1})^{m_{1}}\triangleleft (\partial
^{1})^{n_{1}}(\partial ^{2})^{n_{2}})  \nonumber \\
&=&\delta _{m_{1},n_{1}}\delta
_{m_{2},n_{2}}[[n_{1}]]_{q^{-2}}![[n_{2}]]_{q^{-2}}!,  \nonumber
\end{eqnarray}
which, in turn, leads to 
\begin{eqnarray}
\widetilde{\exp }(\partial _{\hat{R}}\mid x_{L})
&=&\sum_{n_{1},n_{2}=0}^{\infty }\frac{(\partial ^{2})^{n_{2}}(\partial
^{1})^{n_{1}}\otimes (X_{1})^{n_{1}}(X_{2})^{n_{2}}}{%
[[n_{1}]]_{q^{2}}![[n_{2}]]_{q^{2}}!}, \\
\exp (\hat{\partial}_{R}\mid x_{\hat{L}}) &=&\sum_{n_{1},n_{2}=0}^{\infty }%
\frac{(\hat{\partial}^{1})^{n_{1}}(\hat{\partial}^{2})^{n_{2}}\otimes
(X_{2})^{n_{2}}(X_{1})^{n_{1}}}{[[n_{1}]]_{q^{-2}}![[n_{2}]]_{q^{-2}}!},
\end{eqnarray}
where we have introduced coordinates with lower indices by setting 
\begin{equation}
X_{i}=\varepsilon _{ij}X^{j}.
\end{equation}

It is now obvious, from what we have done so far, that the different
exponentials can be transformed into each other by applying some simple
rules. First of all, we can verify the existence of a correspondence given
by 
\begin{eqnarray}
\widetilde{\exp }(x_{\hat{R}} \mid \partial _{L})&\stackrel{{{\QATOP{i}{q}}{%
\QATOP{\rightarrow }{\rightarrow }}{\QATOP{i^{\prime }}{1/q}}}}{%
\longleftrightarrow }&\exp (x_{R}\mid \hat{\partial}_{\hat{L}}), \\
\widetilde{\exp }(\partial _{\hat{R}} \mid x_{L})&\stackrel{{{\QATOP{i}{q}}{%
\QATOP{\rightarrow }{\rightarrow }}{\QATOP{i^{\prime }}{1/q}}}}{%
\longleftrightarrow }&\exp (\hat{\partial}_{R}\mid x_{\hat{L}}),  \nonumber
\end{eqnarray}
where the symbol $\stackrel{{{\QATOP{i}{q}}{\QATOP{\rightarrow }{\rightarrow 
}}{\QATOP{i^{\prime }}{1/q}}}}{\longleftrightarrow }$ indicates a transition
via one of the following two substitutions: 
\begin{eqnarray}
q &\leftrightarrow &q^{-1},\quad \partial _{i}\leftrightarrow \hat{\partial}%
_{i^{\prime }},\quad X^{i}\leftrightarrow X^{i^{\prime }}, \\[0.16in]
q &\leftrightarrow &q^{-1},\quad \partial ^{i}\leftrightarrow \hat{\partial}%
^{i^{\prime }},\quad X_{i}\leftrightarrow X_{i^{\prime }},
\end{eqnarray}
with $i^{^{\prime }}=3-i$. Likewise, one can read off the transformation
rules 
\begin{eqnarray}
\widetilde{\exp }(x_{\hat{R}} \mid \partial _{L}){}&\stackrel{%
i\leftrightarrow i^{\prime }}{\longleftrightarrow }&\widetilde{\exp }%
(\partial _{\hat{R}}\mid x_{L}), \\
\exp (x_{R} \mid \hat{\partial}_{\hat{L}}){}&\stackrel{i\leftrightarrow
i^{\prime }}{\longleftrightarrow }&\exp (\hat{\partial}_{R}\mid x_{\hat{L}}),
\nonumber
\end{eqnarray}
where $\stackrel{i\leftrightarrow i^{\prime }}{\longleftrightarrow }$ now
denotes that one can make a transition between the two expressions by
applying one of the following two substitutions: 
\begin{eqnarray}
X^{i} &\leftrightarrow &\partial ^{i},\quad \partial _{i}\leftrightarrow
X_{i}, \\[0.16in]
X^{i} &\leftrightarrow &\hat{\partial}^{i},\quad \hat{\partial}%
_{i}\leftrightarrow X_{i}.
\end{eqnarray}

\section{3-Dimensional q-deformed Euclidean space\label{Kap2}}

All considerations of the previous section carry over to the q-deformed
Euclidean space with three dimensions\footnote{%
For a definition of 3-dimensional q-deformed Euclidean space see appendix 
\ref{AppQuan} .}. Thus, we limit ourselves to stating the results. As in the
2-dimensional case, there are two different covariant differential calculi
which are completely described by the commutation relations 
\begin{eqnarray}
\partial ^{A}X^{B} &=&g^{AB}+(\hat{R}^{-1})_{CD}^{AB}X^{C}\partial ^{D},
\label{ComRel3dim} \\
\hat{\partial}^{A}X^{B} &=&g^{AB}+(\hat{R})_{CD}^{AB}X^{C}\hat{\partial}%
^{D},\quad A,B\in \{3,+,-\},  \nonumber
\end{eqnarray}
where $\hat{R}$ denotes the R-matrix of the quantum group $SO_{q}(3)$ and $%
g^{AB}$ the corresponding quantum metric. In what follows, we restrict
attention to the first relation in (\ref{ComRel3dim}) from which we have
derived in \cite{BW01} the following expressions: 
\begin{eqnarray}
&&\partial ^{-}\triangleright (X^{+})^{m_{+}}(X^{3})^{m_{3}}(X^{-})^{m_{-}}
\\
&=&-q^{-1}[[m_{+}]]_{q^{4}}(X^{+})^{m_{+}-1}(X^{3})^{m_{3}}(X^{-})^{m_{-}}, 
\nonumber \\[0.16in]
&&\partial ^{3}\triangleright (X^{+})^{m_{+}}(X^{3})^{m_{3}}(X^{-})^{m_{-}}
\\
&=&q^{2m_{+}}[[m_{3}]]_{q^{2}}(X^{+})^{m_{+}-1}(X^{3})^{m_{3}-1}(X^{-})^{m_{-}},
\nonumber \\[0.16in]
&&\partial ^{+}\triangleright (X^{+})^{m_{+}}(X^{3})^{m_{3}}(X^{-})^{m_{-}}
\\
&=&-q^{2m_{3}+1}[[m_{-}]]_{q^{4}}(X^{+})^{m_{+}}(X^{3})^{m_{3}}(X^{-})^{m_{-}-1}
\nonumber \\
&&-\,q\lambda
[[m_{3}]]_{q^{2}}[[m_{3}-1]]_{q^{2}}(X^{+})^{m_{+}+1}(X^{3})^{m_{3}-2}(X^{-})^{m_{-}}.
\nonumber
\end{eqnarray}
Using these formulae we obtain after some tedious
steps 
\begin{eqnarray}
&&\left\langle (\partial ^{+})^{n_{+}}(\partial ^{3})^{n_{3}}(\partial
^{-})^{n_{-}},(X^{+})^{m_{+}}(X^{3})^{m_{3}}(X^{-})^{m_{-}}\right\rangle _{L,%
\hat{R}} \\
&=&\varepsilon \left( (\partial ^{+})^{n_{+}}(\partial
^{3})^{n_{3}}(\partial ^{-})^{n_{-}}\triangleright
(X^{+})^{m_{+}}(X^{3})^{m_{3}}(X^{-})^{m_{-}}\right)   \nonumber \\
&=&\delta _{m_{+},n_{-}}\delta _{m_{3},n_{3}}\delta
_{m_{-},n_{+}}(-q)^{n_{+}-%
\,n_{-}}[[m_{+}]]_{q^{4}}![[m_{3}]]_{q^{2}}![[m_{-}]]_{q^{4}}!.  \nonumber
\end{eqnarray}
Now, we are again in a position to read off the two sets of basis elements
being dual to each other. Finally, this enables us along with 
\begin{equation}
\partial _{A}=g_{AB}\partial ^{B}.
\end{equation}
to write down the exponential as 
\begin{eqnarray}
&&\exp (x_{\hat{R}}\mid \partial _{L})  \label{3dimExp} \\
&=&\sum_{\underline{n}=0}^{\infty }\frac{%
(X^{+})^{n_{+}}(X^{3})^{n_{3}}(X^{-})^{n_{-}}\otimes (\partial
_{-})^{n_{-}}(\partial _{3})^{n_{3}}(\partial _{+})^{n_{+}}}{%
[[n_{+}]]_{q^{4}}![[n_{3}]]_{q^{2}}![[n_{-}]]_{q^{4}}!}.  \nonumber
\end{eqnarray}

Repeating the identical steps as before, we can also compute explicit
formulae for the other types of q-exponentials. These calculations show us
the existence of a correspondence given by 
\begin{eqnarray}
\exp (x_{\hat{R}} \mid \partial _{L})&\stackrel{{\QATOP{\pm }{q}}{\QATOP{%
\rightarrow }{\rightarrow }}{\QATOP{\mp }{1/q}}}{\longleftrightarrow }%
&\widetilde{\exp }(x_{R}\mid \hat{\partial}_{\hat{L}}), \\
\exp (\partial _{\hat{R}} \mid x_{L})&\stackrel{{\QATOP{\pm }{q}}{\QATOP{%
\rightarrow }{\rightarrow }}{\QATOP{\mp }{1/q}}}{\longleftrightarrow }%
&\widetilde{\exp }(\hat{\partial}_{R}\mid x_{\hat{L}}),  \nonumber
\end{eqnarray}
where the symbol $\stackrel{{\QATOP{\pm }{q}}{\QATOP{\rightarrow }{%
\rightarrow }}{\QATOP{\mp }{1/q}}}{\longleftrightarrow }$ indicates a
transition via one of the following two substitutions: 
\begin{eqnarray}
q &\leftrightarrow &q^{-1},\quad \partial _{\pm }\leftrightarrow \hat{%
\partial}_{\mp },\quad \partial _{3}\leftrightarrow \hat{\partial}_{3},\quad
X^{\pm }\leftrightarrow X^{\mp }, \\[0.16in]
q &\leftrightarrow &q^{-1},\quad \partial ^{\pm }\leftrightarrow \hat{%
\partial}^{\mp },\quad \partial ^{3}\leftrightarrow \hat{\partial}^{3},\quad
X_{\pm }\leftrightarrow X_{\mp }.
\end{eqnarray}
Additionally, one can verify the transformation rules 
\begin{eqnarray}
\exp (x_{\hat{R}} \mid \partial _{L}){}&\stackrel{+\leftrightarrow -}{%
\longleftrightarrow }&\exp (\partial _{\hat{R}}\mid x_{L}),  \label{Trans3dim}
\\
\widetilde{\exp }(x_{R} \mid \hat{\partial}_{\hat{L}}){}&\stackrel{%
+\leftrightarrow -}{\longleftrightarrow }&\widetilde{\exp }(\hat{\partial}%
_{R}\mid x_{\hat{L}}),  \nonumber
\end{eqnarray}
where the symbol $\stackrel{+\leftrightarrow -}{\longleftrightarrow }$
denotes that one can make a transition between the two expressions by
applying one of the following two substitutions: 
\begin{eqnarray}
X^{A} &\leftrightarrow &\partial ^{A},\quad \partial _{A}\leftrightarrow
X_{A}, \\[0.16in]
X^{A} &\leftrightarrow &\hat{\partial}^{A},\quad \hat{\partial}%
_{A}\leftrightarrow X_{A}.
\end{eqnarray}

\section{4-Dimensional q-deformed Euclidean space}

The 4-dimensional Euclidean space \cite{Oca96} (for its definition see
appendix \ref{AppQuan}) can be treated along the same line of arguments as
the 2- and 3-dimensional one. Again we begin by considering the commutation
relations between partial derivatives and coordinates, which for the two
covariant differential calculi read 
\begin{eqnarray}
\partial ^{i}X^{j} &=&g^{ij}+q(\hat{R}^{-1})_{kl}^{ij}X^{k}\partial ^{l},
\label{ComR4dim} \\
\hat{\partial}^{i}X^{j} &=&g^{ij}+q^{-1}(\hat{R})_{kl}^{ij}X^{k}\hat{\partial%
}^{l},\quad i,j=1,\ldots ,4,  \nonumber
\end{eqnarray}
with $\hat{R}$ and $g^{ij}$ being the R-matrix of the quantum group $%
SO_{q}(4)$ and the corresponding metric, respectively. From the second
relation in (\ref{ComR4dim}) we have found in \cite{BW01} the following
formulae for the action of partial derivatives on normally ordered
monomials: 
\begin{eqnarray}
\hat{\partial}^{1}\,\bar{\triangleright}\,(X^{1,\ldots ,4})^{\underline{m}}
&=&q^{-1-m_{2}-m_{3}}[[m_{4}]]_{q^{-2}}(X^{1,\ldots ,4})^{\underline{m}%
+(0,0,0,-1)} \\
&&+\,q^{-1}\lambda [[m_{2}]]_{q^{-2}}[[m_{3}]]_{q^{-2}}(X^{1,\ldots ,4})^{%
\underline{m}+(1,-1,-1,0)},  \nonumber \\
\hat{\partial}^{2}\,\bar{\triangleright}\,(X^{1,\ldots ,4})^{\underline{m}}
&=&q^{-m_{1}}[[m_{3}]]_{q^{-2}}(X^{1,\ldots ,4})^{\underline{m}+(0,0,-1,0)} 
\nonumber \\
\hat{\partial}^{3}\,\bar{\triangleright}\,(X^{1,\ldots ,4})^{\underline{m}}
&=&q^{-m_{1}}[[m_{2}]]_{q^{-2}}(X^{1,\ldots ,4})^{\underline{m}+(0,-1,0,0)},
\nonumber \\
\hat{\partial}^{4}\,\bar{\triangleright}\,(X^{1,\ldots ,4})^{\underline{m}}
&=&q[[m_{1}]]_{q^{-2}}(X^{1,\ldots ,4})^{\underline{m}+(-1,0,0,0)}, 
\nonumber
\end{eqnarray}
where, for compactness, we have introduced a new notation: 
\begin{eqnarray}
(X^{1,\ldots ,4})^{\underline{m}}
&=&(X^{1})^{m_{1}}(X^{2})^{m_{2}}(X^{3})^{m_{3}}(X^{4})^{m_{4}}, \\
(\hat{\partial}_{4,\ldots ,1})^{\underline{n}} &=&(\hat{\partial}%
_{4})^{n_{4}}(\hat{\partial}_{3})^{n_{3}}(\hat{\partial}_{2})^{n_{2}}(\hat{%
\partial}_{1})^{n_{1}}.  \nonumber
\end{eqnarray}
Switching to partial derivatives with lower indices, 
\begin{equation}
\hat{\partial}_{i}=g_{ij}\hat{\partial}^{j},
\end{equation}
these expressions imply for the dual pairing the identity 
\begin{eqnarray}
&&\left\langle (\hat{\partial}_{4,\ldots ,1})^{\underline{n}},(X^{1,\ldots
,4})^{\underline{m}}\right\rangle _{\hat{L},R} \\
&=&\delta _{m_{1},n_{1}}\delta _{m_{2},n_{2}}\delta _{m_{3},n_{3}}\delta
_{m_{4},n_{4}}  \nonumber \\
&&\cdot
\,[[m_{1}]]_{q^{-2}}![[m_{2}]]_{q^{-2}}![[m_{3}]]_{q^{-2}}![[m_{4}]]_{q^{-2}}!,
\nonumber
\end{eqnarray}
which, with the same reasonings as in the previous sections, leads to 
\begin{equation}
\exp (x_{R}\mid \hat{\partial}_{\hat{L}})=\sum_{\underline{n}=0}^{\infty }%
\frac{(X^{1,\ldots ,4})^{\underline{n}}\otimes (\hat{\partial}_{4,\ldots
,1})^{\underline{n}}}{%
[[n_{1}]]_{q^{-2}}![[n_{2}]]_{q^{-2}}![[n_{3}]]_{q^{-2}}![[n_{4}]]_{q^{-2}}!}%
.  \label{qExp4}
\end{equation}

In complete analogy to the 2- and 3-dimensional case there is again
a correspondence between the different types of q-exponentials. 
First of all, we have 
\begin{eqnarray}
\exp (x_{R} \mid \hat{\partial}_{\hat{L}}){}&\stackrel{{{\QATOP{i}{q}}{%
\QATOP{\rightarrow }{\rightarrow }}{\QATOP{i^{\prime }}{1/q}}}}{%
\longleftrightarrow }&\widetilde{\exp }(x_{\hat{R}}\mid \partial _{L}),
\label{TransFor4dim} \\
\widetilde{\exp }(\partial _{\hat{R}} \mid x_{L}){}&\stackrel{{{\QATOP{i}{q}%
}{\QATOP{\rightarrow }{\rightarrow }}{\QATOP{i^{\prime }}{1/q}}}}{%
\longleftrightarrow }&\exp (\hat{\partial}_{R}\mid x_{\hat{L}}),  \nonumber
\end{eqnarray}
which concretely means that the expressions on the right- and left-hand side
can be transformed into each other by one of the following two
substitutions: 
\begin{eqnarray}
q &\leftrightarrow &q^{-1},\quad \partial _{i}\leftrightarrow \hat{\partial}%
_{i^{\prime }},\quad X^{i}\leftrightarrow X^{i^{\prime }}, \\[0.16in]
q &\leftrightarrow &q^{-1},\quad \partial ^{i}\leftrightarrow \hat{\partial}%
^{i^{\prime }},\quad X_{i}\leftrightarrow X_{i^{\prime }},
\end{eqnarray}
where $i=1,\ldots ,4,$ and $i^{^{\prime }}=5-i.$ In complete analogy to the
2- and 3-dimensional case we can also find the transformations 
\begin{eqnarray}
\widetilde{\exp }(x_{\hat{R}} \mid \partial _{L}){}&\stackrel{%
i\leftrightarrow i^{\prime }}{\longleftrightarrow }&\widetilde{\exp }%
(\partial _{\hat{R}}\mid x_{L}), \\
\exp (x_{R} \mid \hat{\partial}_{\hat{L}}){}&\stackrel{i\leftrightarrow
i^{\prime }}{\longleftrightarrow }&\exp (\hat{\partial}_{R}\mid x_{\hat{L}}) 
\nonumber
\end{eqnarray}
symbolizing a transition via one of the following two substitutions: 
\begin{eqnarray}
X^{i} &\leftrightarrow &\partial ^{i},\quad \partial _{i}\leftrightarrow
X_{i}, \\[0.16in]
X^{i} &\leftrightarrow &\hat{\partial}^{i},\quad \hat{\partial}%
_{i}\leftrightarrow X_{i}.
\end{eqnarray}

\section{q-Deformed Minkowski space}

From a physical point of view the most important case we want to discuss in
this article is q-deformed Minkowski space \cite{CSSW90,SWZ91,Maj91}
\footnote{
For its definition see appendix \ref{AppQuan}. Other versions of
q-deformed Minkowski space are given in \cite{Dob94,ChDe95}.}.
There are again two
covariant differential calculi given by 
\begin{eqnarray}
\partial ^{\mu }X^{\nu } &=&g^{\mu \nu }+q^{-2}(\hat{R}_{II}^{-1})_{\rho
\sigma }^{\mu \nu }X^{\rho }\partial ^{\sigma }, \\
\hat{\partial}^{\mu }X^{\nu } &=&g^{\mu \nu }+q^{2}(\hat{R}_{II})_{\rho
\sigma }^{\mu \nu }X^{\rho }\hat{\partial}^{\sigma },
\quad \mu ,\nu \in \{\pm,0,3\},  \nonumber
\end{eqnarray}
where $\hat{R}_{II}$ stands for one of the two R-matrices of the q-deformed
Lorentz-algebra \cite{LSW94} and $g^{\mu \nu }$ for the corresponding
quantum metric. From the above relations we have calculated in \cite{BW01}
representations for partial derivatives. However, the complexity of these
representations makes it rather difficult to deduce for the dual pairing a
closed expression from which we could read off the two sets of basis
elements. Thus, we cannot directly apply the procedure of the last two
sections for determining a basis being dual to a given one of normally ordered
monomials. For this reason we would like to present a different method for
calculating q-exponentials.

To begin, our first job is now to seek a useful ansatz describing the
q-exponentials. Since our exponentials are required to be bosonic \cite{KM94}
they have to satisfy the properties 
\begin{eqnarray}
&&\Lambda \triangleright \Big (\exp (x_{R}\mid \hat{\partial}_{\hat{L}})\Big
)  \label{ExpLam} \\
&=&(\Lambda \otimes \Lambda )\triangleright \exp (x_{R}\mid \hat{\partial}_{%
\hat{L}})  \nonumber \\
&=&\varepsilon (\Lambda )\exp (x_{R}\mid \hat{\partial}_{\hat{L}})=\exp
(x_{R}\mid \hat{\partial}_{\hat{L}}),  \nonumber \\[0.16in]
&&\tau ^{3}\triangleright \Big (\exp (x_{R}\mid \hat{\partial}_{\hat{L}})%
\Big )  \label{ExpTau} \\
&=&(\tau ^{3}\otimes \tau ^{3})\triangleright \exp (x_{R}\mid \hat{\partial}%
_{\hat{L}})  \nonumber \\
&=&\varepsilon (\tau ^{3})\exp (x_{R}\mid \hat{\partial}_{\hat{L}})=\exp
(x_{R}\mid \hat{\partial}_{\hat{L}}),  \nonumber
\end{eqnarray}
with $\tau ^{3}$ being a grouplike generator of the q-Lorentz algebra and $%
\Lambda $ denoting the associated scaling operator \cite{OSWZ92}. Recalling
that 
\begin{eqnarray}
\Lambda \triangleright X^{\mu } &=&q^{-2}X^{\mu }, \\
\Lambda \triangleright \hat{\partial}^{\mu } &=&q^{2}\hat{\partial}^{\mu
},\qquad \mu \in \{+,3,0,-\},  \nonumber
\end{eqnarray}
and 
\begin{eqnarray}
\tau ^{3}\triangleright X^{\pm } &=&q^{\mp 4}X^{\pm },\quad \tau
^{3}\triangleright X^{0}=X^{0},\quad \tau ^{3}\triangleright X^{3/0}=X^{3/0},
\\
\tau ^{3}\triangleright \hat{\partial}^{\pm } &=&q^{\mp 4}\hat{\partial}%
^{\pm },\quad \tau ^{3}\triangleright \hat{\partial}^{0}=\hat{\partial}%
^{0},\quad \tau ^{3}\triangleright \hat{\partial}^{3/0}=\hat{\partial}^{3/0},
\nonumber
\end{eqnarray}
together with the identities in (\ref{ExpLam}) and (\ref{ExpTau})
establishes that the exponentials have to take the form\footnote{%
For notational convenience, we introduce a multi-index \underline{$m$}$%
\,\equiv (m_{+},m_{3/0},m_{3},m_{-})$ .} 
\begin{equation}
\exp (x_{R}\mid \hat{\partial}_{\hat{L}})=\sum_{\underline{m}=0}^{\infty }f^{%
\underline{m}}(\underline{X})\otimes (\hat{\partial}^{+})^{m_{-}}(\hat{%
\partial}^{3/0})^{m_{3}}(\hat{\partial}^{0})^{m_{3/0}}(\hat{\partial}%
^{-})^{m_{+}},  \label{ExpMinkAns}
\end{equation}
where 
\begin{eqnarray}
&&f^{\underline{m}}(\underline{X})  \label{KoefExpMin} \\
&=&\sum_{\QATOP{2l+v\leq m_{3}}{-m_{3/0}\leq v,\,-m_{\pm }\leq l}}f_{l,v}^{%
\underline{m}}\cdot
(X^{+})^{m_{+}+\,l}(X^{3/0})^{m_{3/0}+v}(X^{3})^{m_{3}-2l-v}(X^{-})^{m_{-}+%
\,l}.  \nonumber
\end{eqnarray}

In the following it is our aim to determine the unknown coefficients $%
f_{l,v}^{\underline{m}}$. Before doing this let us introduce, for brevity, 
\begin{eqnarray}
\hat{\partial}^{\,\underline{k}} &\equiv &(\hat{\partial}^{+})^{k_{-}}(\hat{%
\partial}^{3/0})^{k_{3}}(\hat{\partial}^{0})^{k_{3/0}}(\hat{\partial}%
^{-})^{k_{+}}, \\
X^{\underline{k}} &\equiv
&(X^{+})^{k_{+}}(X^{3/0})^{k_{3/0}}(X^{3})^{k_{3}}(X^{-})^{k_{-}}.  \nonumber
\end{eqnarray}
Inserting the expressions of (\ref{ExpMinkAns}) and (\ref{KoefExpMin}) into%
\footnote{%
This formula follows from a direct application of (\ref{NorEbWel}) and (\ref
{EigGl}).} 
\begin{equation}
(\varepsilon \otimes id)\circ (\hat{\partial}^{\,\underline{k}}\otimes
id)\triangleright \exp (x_{R}\mid \hat{\partial}_{\hat{L}})=\hat{\partial}%
^{\,\underline{k}}
\end{equation}
provides us with a system of equations given by 
\begin{equation}
\sum_{\QATOP{2l+v\leq m_{3}}{-m_{3/0}\leq v,{}-\min (m_{+},m_{-})\leq l}%
}f_{l,v}^{\underline{m}}\cdot \left\langle \hat{\partial}^{\,\underline{k}%
},X^{\underline{m}+(l,v,-2l-v,l)}\right\rangle _{\hat{L},R}=\delta _{%
\underline{m},\underline{k}},  \label{GlSysExpMin}
\end{equation}
where 
\begin{equation}
\delta _{\underline{m},\underline{k}}=\delta _{m_{+},k_{+}}\delta
_{m_{3},k_{3}}\delta _{m_{3/0},k_{3/0}}\delta _{m_{-},k_{-}}.
\end{equation}
This system for the unknown coefficients $f_{l,v}^{\underline{m}}$ can be
simplified further by taking the relations 
\begin{equation}
\left\langle \hat{\partial}^{\,\underline{m}},X^{\underline{m}%
+(l,v,-2l-v,l)}\right\rangle _{\hat{L},R}=0,\quad \text{if }2l+v>0\text{ or }%
v>0,  \label{EigDualPaar}
\end{equation}
a proof of which is given in appendix \ref{Proofs}. By exploiting the
property (\ref{EigDualPaar}) one can then show that we have 
\begin{equation}
f_{l,v}^{\underline{m}}=0,\quad \text{if }v<0\text{ or }2l+v<0.
\label{Propf}
\end{equation}
The proof of this assumption can again be found in appendix \ref{Proofs}.
Finally, a little thought using (\ref{EigDualPaar}) and (\ref{Propf}) shows
that the system (\ref{GlSysExpMin}) can be reduced to 
\begin{equation}
\sum_{\QATOP{0\leq 2l+v\leq m_3}{0\leq v
,-\min (m_{+},m_{-})\leq l}}f_{l,v}^{\underline{m}}\cdot \left\langle \hat{%
\partial}^{\,\underline{m}+(l^{\prime },v^{\prime },-2l^{\prime }-v^{\prime
},l^{\prime })},X^{\underline{m}+(l,v,-2l-v,l)}\right\rangle _{\hat{L}%
,R}=\delta _{0}^{v^{\prime }}\delta _{0}^{l^{\prime }},  \label{GlSysneu}
\end{equation}
if \underline{$k$} is specified according to 
\begin{eqnarray}
\underline{k} &=&\underline{m}+(l^{\prime },v^{\prime },-2l^{\prime
}-v^{\prime },l^{\prime }) \\
&=&(m_{+}+l^{\prime },m_{3/0}+v^{\prime },m_{3}-2l^{\prime }-v^{\prime
},m_{-}+l^{\prime }),  \nonumber
\end{eqnarray}
where $l^{\prime }$ and $v^{\prime }$ are non-negative integers with 
\begin{equation}
2l^{\prime }+v^{\prime }\leq m_{3}.
\end{equation}

It is our next goal to present a method for solving the above system of
equations. Towards this end we introduce the function 
\begin{equation}
z(v,l)\equiv v+l+\Big[\frac{v}{2}\Big]+1+\sum_{i=0}^{v-1}\Big(\Big[\frac{%
m_{3}-i}{2}\Big]+\Big[\frac{i}{2}\Big]\Big),
\end{equation}
where $[s]$ denotes the biggest integer not being bigger than $s$. From the
constraints on the summations in (\ref{GlSysExpMin}) and (\ref{GlSysneu}) we
know that the integer values the variables $v$ and $l$ shall take on are
restricted to 
\begin{eqnarray}
0 &\leq &v\leq v_{\max }(\underline{m}), \\
-\Big[\frac{v}{2}\Big]  &\leq &l\leq \Big[\frac{m_{3}-v}{2}\Big],  \nonumber
\end{eqnarray}
where we have set $v_{\max }(\underline{m})\equiv m_{3}+2\min (m_{+},m_{-}).$
For a better understanding of the following considerations it is important
to notice that $z(v,l)$ shows the property 
\begin{equation}
z(v,l)<z(v^{\prime },l^{\prime })\Leftrightarrow
 \left\{\begin{array}{l} 
 v<v^{\prime},\\v=v^{\prime },\quad l<l^{\prime }. \end{array} \right.
\end{equation}
This implies that the maximum value of $z(v,l)$ is given by 
\begin{eqnarray}
z_{\max }(\underline{m})&\equiv& v_{\max }
(\underline{m})-\min(m_+,m_-)+\Big[\frac{v_{\max }(%
\underline{m})}{2}\Big]+1\\ \nonumber
&&+\sum_{i=0}^{v_{\max }(\underline{m})}\Big(\Big[%
\frac{m_{3}-i}{2}\Big]+\Big[\frac{i}{2}\Big]\Big).
\end{eqnarray}
It is not very difficult to convince oneself that we can also establish a
one-to-one correspondence between the allowed values of $v$ and $l$ on the
one hand and those of $z(v,l)$ on the other hand by setting 
\begin{eqnarray}
v_{z}(m_{3}) &\equiv &\max \Big \{j\in \mathbb{N}_{0}\mid z-j-\sum_{i=0}^{j-1}%
\Big(\Big[\frac{m_{3}-i}{2}\Big]+\Big[\frac{i}{2}\Big]\Big)>0\Big \}, \\
l_{z}(m_{3}) &\equiv &z-v_{z}(m_{3})-\Big[\frac{v_{z}(m_{3})}{2}\Big]%
-1-\sum_{i=0}^{v_{z}(m_{3})-1}\Big(\Big[\frac{m_{3}-i}{2}\Big]+\Big[\frac{i}{%
2}\Big]\Big).  \nonumber
\end{eqnarray}
The deeper reason for introducing the function $z(v,l)$ becomes quite clear,
as soon as one realizes that it establishes an ordering for the coefficients 
$f_{l,v}^{\underline{m}}$ if we take the convention 
\begin{equation}
f_{z(v,l)}^{\underline{m}}\equiv f_{v,l}^{\underline{m}}.
\end{equation}
By using this ordering the system (\ref{GlSysneu}) can be rewritten as 
\begin{equation}
\sum_{j=1}^{z_{\max }(\underline{m})}\Theta \left( l_{j}+\min
(m_{+},m_{-})\right) \cdot f_{j}^{\underline{m}}\cdot \left\langle
z(v^{\prime },l^{\prime }),j\right\rangle_{\hat{L},R}^{\underline{m}}=\delta
_{0}^{v^{\prime }}\delta _{0}^{l^{\prime }}, \label{GlUm}
\end{equation}
where we have introduced the step-function 
\begin{equation}
\Theta (h)=\left\{ \begin{array}{r@{\quad,\quad}l}
0& \text{if }h<0\\1&\text{otherwise}\end{array} \right.
\end{equation}
and as a shorthand notation\footnote{To understand the equivalence of 
(\ref{GlSysneu}) and (\ref{GlUm}) it is helpful to keep in mind that 
$v_{z(v^{\prime },l^{\prime })}=v^{\prime }$ and 
$l_{z(v^{\prime },l^{\prime })}=l^{\prime }$.} 
\begin{equation}
\left\langle k,j\right\rangle _{\hat{L},R}^{\underline{m}}\equiv
\left\langle \hat{\partial}^{\,\underline{m}+(l_{k},v_{k},-2l_{k}-%
\,v_{k},l_{k})},X^{\underline{m}+(l_{j},v_{j},-2l_{j}-v_{j},l_{j})}\right%
\rangle _{\hat{L},R}.  \label{AbkDualPaar}
\end{equation}
Generalizing relation (\ref{EigDualPaar}) to\footnote{%
For a proof of this assumption see appendix \ref{Proofs}.} 
\begin{equation}
\left\langle k,j\right\rangle _{\hat{L},R}^{\underline{m}}=0,\quad \text{%
if\quad }k<j,  \label{VerAlg}
\end{equation}
shows us that it is sufficient to choose $z(v^{\prime },l^{\prime })$ as an
upper bound of the sum over $j$, i.e. 
\begin{equation}
\sum_{j=1}^{z(v^{\prime },l^{\prime })}\Theta \left( l_{j}+\min
(m_{+},m_{-})\right) \cdot f_{j}^{\underline{m}}\cdot \left\langle
z(v^{\prime },l^{\prime }),j\right\rangle_{\hat{L},R}^{\underline{m}}=\delta
_{0}^{v^{\prime }}\delta _{0}^{l^{\prime }}
\end{equation}
or 
\begin{equation}
\sum_{j=1}^{k}\Theta \left( l_{j}+\min (m_{+},m_{-})\right) \cdot f_{j}^{%
\underline{m}}\cdot \left\langle
k,j\right\rangle_{\hat{L},R}^{\underline{m}}
=\delta
_{0}^{v_{k}}\delta _{0}^{l_{k}},
\end{equation}
where $1\leq k\leq z_{\max }(\underline{m}).$ In this way we have arrived at
a system of triangular form which we can reduce to the recursion relation 
\begin{eqnarray}
f_{k}^{\underline{m}} &=&-\sum_{1\leq j<k}\Theta \left( l_{j}+\min
(m_{+},m_{-})\right)  \\[0.16in]
&\times &f_{j}^{\underline{m}}\cdot \frac{\left\langle k,j\right\rangle _{%
\hat{L},R}^{\underline{m}}}{\left\langle k,k\right\rangle _{\hat{L},R}^{%
\underline{m}}},\quad \text{for }1<k\leq z_{\max }(\underline{m}),  \nonumber
\\
f_{1}^{\underline{m}} &=&\frac{1}{\left\langle 1,1\right\rangle _{\hat{L}%
,R}^{\underline{m}}}=\langle \hat{\partial}^{\,\underline{m}},X^{\underline{m%
}}\rangle _{\hat{L},R}  \label{fm1} \\
&=&\frac{(-q)^{m_{-}-\,m_{+}}}{%
[[m_{-}]]_{q^{2}}![[m_{3/0}]]_{q^{2}}![[m_{3}]]_{q^{2}}![[m_{+}]]_{q^{2}}!}.
\nonumber
\end{eqnarray}
Notice that the last expression for $f_{1}^{\underline{m}}$ in Eqn.(\ref{fm1}%
) can be derived rather easily by multiple application of the
representations presented in \cite{BW01}. Now, it should be obvious, that
the coefficients $f_{k}^{\underline{m}}$ can be expressed as 
\begin{eqnarray}
f_{k}^{\underline{m}} &=&\sum_{i=1}^{k-1}\sum_{1=j_{0}<j_{1}<\ldots
<j_{i}=k}(-1)^{i} \\
&\times &\frac{\prod_{p=1}^{i}\Theta \left( l_{j_{p}}+\min
(m_{+},m_{-})\right) \cdot \left\langle j_{p},j_{p-1}\right\rangle _{\hat{L%
},R}^{\underline{m}}}{\prod_{r=0}^{i}\left\langle j_{r},j_{r}\right\rangle _{%
\hat{L},R}^{\underline{m}}}.  \nonumber
\end{eqnarray}
Substituting this into Eqn.(\ref{KoefExpMin}) together with (\ref{ExpMinkAns}%
) finally yields 
\begin{eqnarray}
&&\exp (x_{R}\mid \hat{\partial}_{\hat{L}})  \label{ExpMin} \\
&=&\sum_{\underline{m}=0}^{\infty }\sum_{k=1}^{z_{\max}(\underline{m})}
(C)_{k}^{%
\underline{m}}\cdot \frac{(-q)^{m_{-}-\,m_{+}}\cdot X^{\underline{m}%
+(l_{k},v_{k},-2l_{k}-v_{k},l_{k})}\otimes \hat{\partial}^{\,\underline{m}}}{%
[[m_{-}]]_{q^{2}}![[m_{3/0}]]_{q^{2}}![[m_{3}]]_{q^{2}}![[m_{+}]]_{q^{2}}!},
\nonumber
\end{eqnarray}
where 
\begin{eqnarray}
(C)_{k}^{\underline{m}} &=&\sum_{i=1}^{k-1}{}\sum_{1=j_{0}<j_{1}<\ldots
<j_{i}=k}(-1)^{i} \\
&\times &\prod_{p=1}^{i}\frac{\Theta \left( l_{j_{p}}+\min
(m_{+},m_{-})\right) \cdot \left\langle j_{p},j_{p-1}\right\rangle _{\hat{L%
},R}^{\underline{m}}}{\left\langle j_{r},j_{r}\right\rangle _{\hat{L},R}^{%
\underline{m}}}.  \nonumber
\end{eqnarray}
Taking into account the conjugation properties \cite{Maj94star} 
\begin{equation}
\overline{\sum_a f^a\otimes e_a}=\sum_a \overline{e_a}\otimes\overline{f^a}
\end{equation}
and 
\begin{eqnarray}
\overline{X^{0}} &=&X^{0},\quad \overline{X^{3/0}}=X^{3/0},\quad \overline{%
X^{\pm }}=-q^{\mp 1}X^{\mp }, \\
\overline{\hat{\partial}^{0}} &=&\hat{\partial}^{0},\quad \overline{\hat{%
\partial}^{3/0}}=\hat{\partial}^{3/0},\quad \overline{\hat{\partial}^{\pm }}%
=-q^{\mp 1}\hat{\partial}^{\mp },  \nonumber
\end{eqnarray}
immediately gives us 
\begin{eqnarray}
&&\exp (\hat{\partial}_{R}\mid x_{\hat{L}})=
\overline{\exp (x_{R}\mid \hat{\partial}_{\hat{L}})} \\
&=&\sum_{\underline{m}=0}^{\infty }\sum_{k=1}^{z_{\max }(m_{3})}(\bar{C}%
)_{k}^{\underline{m}}\cdot \frac{(-q)^{m_{+}-\,m_{-}}\cdot \hat{\partial}^{%
\underline{\,m}}\otimes X^{\underline{m}+(l_{k},v_{k},-2l_{k}-v_{k},l_{k})}}{%
[[m_{-}]]_{q^{2}}![[m_{3/0}]]_{q^{2}}![[m_{3}]]_{q^{2}}![[m_{+}]]_{q^{2}}!},
\nonumber
\end{eqnarray}
where 
\begin{eqnarray}
(\bar{C})_{k}^{\underline{m}}
&=&\sum_{i=1}^{k-1}{}\sum_{1=j_{0}<j_{1}<\ldots <j_{i}=k}(-1)^{i} \\
&\times &\prod_{p=1}^{i}\frac{\Theta \left( l_{j_{p}}+\min
(m_{+},m_{-})\right) \overline{\left\langle j_{p-1},j_{p}\right\rangle }_{%
\hat{L},R}^{\,\underline{m}}}{\overline{\left\langle
j_{r},j_{r}\right\rangle }_{\hat{L},R}^{\,\underline{m}}}  \nonumber
\end{eqnarray}
with 
\begin{equation}
\overline{\left\langle k,j\right\rangle }_{\hat{L},R}^{\,\underline{m}%
}\equiv \left\langle X^{\,\underline{m}+(l_{j},v_{j},-2l_{j}-v_{j},l_{j})},%
\hat{\partial}^{\,\underline{m}+(l_{k},v_{k},-2l_{k}-\,v_{k},l_{k})}\right%
\rangle _{\hat{L},R}.  \label{AbkDuP2}
\end{equation}
In this sense, we have found expressions for q-exponentials in terms of the
dual pairing between coordinates and derivatives. It remains to evaluate the
expressions in (\ref{AbkDualPaar}) and (\ref{AbkDuP2}). But this is a rather
tedious task which has to be done elsewhere. Thus, we do not want to discuss
that issue any further here.

For completeness we wish to present the rules making a connection between
the different types of q-exponentials. With the same reasonings already
applied to the Euclidean cases we can now write 
\begin{eqnarray}
\exp (x_{R} \mid \hat{\partial}_{\hat{L}}){}&\stackrel{{{\QATOP{\pm }{q}}{%
\QATOP{\rightarrow }{\rightarrow }}{\QATOP{\mp }{1/q}}}}{\longleftrightarrow 
}&\widetilde{\exp }(x_{\hat{R}}\mid \partial _{L}), \\
\exp (\hat{\partial}_{R} \mid x_{\hat{L}}){}&\stackrel{{{\QATOP{\pm }{q}}{%
\QATOP{\rightarrow }{\rightarrow }}{\QATOP{\mp }{1/q}}}}{\longleftrightarrow 
}&\widetilde{\exp }(\partial _{\hat{R}}\mid x_{L}),  \nonumber
\end{eqnarray}
where $\stackrel{{{\QATOP{\pm }{q}}{\QATOP{\rightarrow }{\rightarrow }}{%
\QATOP{\mp }{1/q}}}}{\longleftrightarrow }$ symbolizes the substitutions 
\begin{eqnarray}
q &\leftrightarrow& q^{-1},\quad X^{\pm }\leftrightarrow X^{\mp },\\
\hat{\partial}^{\pm }&\leftrightarrow&
\partial ^{\mp },\quad \hat{\partial}^{3/0}\leftrightarrow \partial
^{3/0},\quad \hat{\partial}^{0}\leftrightarrow \partial
^{0}.\nonumber
\end{eqnarray}

\section{Remarks}

Let us end with a few comments on the explicit expressions we have
derived for
q-exponentials. In the Euclidean cases the formulae for dual pairing and
q-exponential are in complete analogy to those of their classical
counterparts. In particular, they do not depend on terms proportional to
powers of the deformation parameter $\lambda =q-q^{-1}$. In the case of
q-deformed Minkowski space the situation is a little bit different, as it is
impossible to find two sets of normally ordered monomials which constitute
two bases being dual to each other. In other words, for every choice of
normally ordered monomials there are terms like 
\begin{equation}
\left\langle \hat{\partial}^{-}\hat{\partial}^{+},(X^{3})^{2}\right\rangle _{%
\hat{L},R}=\lambda (1+q\lambda _{+}^{-1}),\quad \lambda _{+}=q+q^{-1},
\end{equation}
vanishing in the undeformed limit as $q\rightarrow 1.$ It is for this reason
that in expression (\ref{ExpMin}) non-classical factors $(C)_{k}^{\underline{%
m}}$ appear which depend on powers of $\lambda $ in such a way that 
\begin{equation}
(C)_{k}^{\underline{m}}\stackrel{q\rightarrow 1}{\longrightarrow } \left\{
\begin{array}{r@{\quad,\quad}l} 1& \text{if }k=1 \\
0&\text{otherwise} \end{array} \right.
\end{equation}

\noindent \textbf{Acknowledgement}\newline
First of all I want to express my gratitude to Julius Wess for his efforts,
suggestions and discussions. Also I would like to thank Fabian Bachmaier,
Dzo Mikulovic, Alexander Schmidt and Michael Wohlgenannt for useful
discussions and their steady support. Finally, I would like to thank
the referees for their helpful comments.

\appendix

\section{Quantum spaces\label{AppQuan}}

In this appendix we list for the quantum spaces under
consideration the explicit form of their defining commutation relations,
their conjugation properties and the nonvanishing elements of the quantum
metric.

The coordinates of 2-dimensional q-deformed Euclidean space fulfil the
relation 
\begin{equation}
X^{1}X^{2}=qX^{2}X^{1},  \label{2dimQuan}
\end{equation}
whereas the quantum metric is given by a matrix $\varepsilon ^{ij}$ with
non-vanishing elements 
\begin{equation}
\varepsilon ^{12}=q^{-1/2},\quad \varepsilon ^{21}=-q^{1/2}.
\end{equation}
Furthermore, the relation (\ref{2dimQuan}) is compatible with the
conjugation assignment: 
\begin{equation}
\overline{X^{i}}=-\varepsilon _{ij}X^{j}
\end{equation}
where $\varepsilon _{ij}$ denotes the inverse of $\varepsilon ^{ij}.$

In the case of q-deformed Euclidean space in three dimensions the
commutation relations read 
\begin{eqnarray}
X^{3}X^{+} &=&q^{2}X^{+}X^{3},  \label{Koord3dim} \\
X^{-}X^{3} &=&q^{2}X^{3}X^{-},  \nonumber \\
X^{-}X^{+} &=&X^{+}X^{-}+\lambda X^{3}X^{3}.  \nonumber
\end{eqnarray}
The non-vanishing elements of the quantum metric are 
\begin{equation}
g^{+-}=-q,\quad g^{33}=1,\quad g^{-+}=-q^{-1}.
\end{equation}
Now, the conjugation of the coordinates is given by 
\begin{equation}
\overline{X^{A}}=g_{AB}X^{B}  \label{KonRel}
\end{equation}
with $g_{AB}$ being the inverse of $g^{AB}.$

For the 4-dimensional Euclidean space, we have the relations 
\begin{eqnarray}
X^{1}X^{2} &=&qX^{2}X^{1},  \label{Algebra4} \\
X^{1}X^{3} &=&qX^{3}X^{1},  \nonumber \\
X^{3}X^{4} &=&qX^{4}X^{3},  \nonumber \\
X^{2}X^{4} &=&qX^{4}X^{2},  \nonumber \\
X^{2}X^{3} &=&X^{3}X^{2},  \nonumber \\
X^{4}X^{1} &=&X^{1}X^{4}+\lambda X^{2}X^{3}.
\end{eqnarray}
The metric has the non-vanishing components 
\begin{equation}
g^{14}=q^{-1},\quad g^{23}=g^{32}=1,\quad g^{41}=q.
\end{equation}
The inverse $g_{ij}$ of this metric can again be used to formulate the
conjugation properties for the coordinates, i.e. 
\begin{equation}
\overline{X^{i}}=g_{ij}X^{j}.  \label{Metrik}
\end{equation}

For q-deformed Minkowski space one has the relations 
\begin{eqnarray}
X^{\mu }X^{0} &=&X^{0}X^{\mu },\quad \mu \in \{0,+,-,3\},  \label{Minrel} \\
X^{-}X^{3}-q^{2}X^{3}X^{-} &=&-q\lambda X^{0}X^{-},  \nonumber \\
X^{3}X^{+}-q^{2}X^{+}X^{3} &=&-q\lambda X^{0}X^{+},  \nonumber \\
X^{-}X^{+}-X^{+}X^{-} &=&\lambda (X^{3}X^{3}-X^{0}X^{3}),  \nonumber
\end{eqnarray}
the metric 
\begin{equation}
g^{00}=-1,\quad g^{33}=1,\quad g^{+-}=-q,\quad g^{-+}=-q^{-1}.
\end{equation}
Finally, the conjugation on q-deformed Minkowski space is determined by 
\begin{equation}
\overline{X^{0}}=X^{0},\quad \overline{X^{3}}=X^{3},\quad \overline{X^{\pm }}%
=-q^{\mp 1}X^{\mp }.
\end{equation}

\section{Proofs\label{Proofs}}

\begin{enumerate}
\item  \textbf{The Proof of (\ref{EigDualPaar}) and (\ref{VerAlg}):}\newline
Recalling that $\hat{\partial}^{3/0}$ obeys the commutation relations 
\begin{equation}
\hat{\partial}^{3/0}\hat{\partial}^{0}=\hat{\partial}^{0}\hat{\partial}%
^{3/0},\quad \hat{\partial}^{3/0}\hat{\partial}^{-}=q^{-2}\hat{\partial}^{-}%
\hat{\partial}^{3/0}
\end{equation}
and 
\begin{equation}
\hat{\partial}^{3/0}X^{+}=X^{+}\hat{\partial}^{3/0},\quad \hat{\partial}%
^{3/0}X^{3/0}=X^{3/0}\hat{\partial}^{3/0},
\end{equation}
we can rewrite the dual pairing as follows: 
\begin{eqnarray}
&&\left\langle \hat{\partial}^{\,\underline{m}},X^{\underline{m}%
+(l,v,-2l-v,l)}\right\rangle _{\hat{L},R}  \label{UmfDP} \\
&=&q^{2lm_{3}}\Big [(\hat{\partial}^{+})^{m_{-}}(\hat{\partial}%
^{0})^{m_{3/0}}(\hat{\partial}^{-})^{m_{+}}\,\bar{\triangleright}  \nonumber
\\
&&\left. (X^{+})^{m_{+}+l}(X^{3})^{m_{3/0}+v}((\hat{\partial}%
^{3/0})^{m_{3}}\,\bar{\triangleright}\,(X^{3})^{m_{3}-2l-v}(X^{-})^{m_{-}+l})%
\Big ]\right| _{\underline{X}=0}.  \nonumber
\end{eqnarray}
Direct inspection of the representations of $\hat{\partial}^{3/0}$
(presented in \cite{BW01}) shows that 
\begin{equation}
(\hat{\partial}^{3/0})^{n}\,\bar{\triangleright}%
\,(X^{3})^{k_{3}}(X^{-})^{k_{-}}=0,\quad \text{if }n>k_{3},
\label{PropPar3/0}
\end{equation}
which, in turn, tells us that the last expression in (\ref{UmfDP}) has to
vanish, if $2l+v>0.$\newline
Now we come to the case $v>0.$ First of all, let us note that the roles of
coordinates and derivatives can be completely reversed. In this sense we
proceed as follows: 
\begin{eqnarray}
&&\left\langle \hat{\partial}^{\,\underline{m}},X^{\underline{m}%
+(l,v,-2l-v,l)}\right\rangle _{\hat{L},R}  \label{UmfDulPaar} \\
&=&\left. (\hat{\partial}^{\,\underline{m}}\,\bar{\triangleright}\,X^{%
\underline{m}+(l,v,-2l-v,l)})\right| _{\underline{X}=0}  \nonumber \\
&=&\left. (\hat{\partial}^{\,\underline{m}}\triangleleft X^{\underline{m}%
+(l,v,-2l-v,l)})\right| _{\underline{\hat{\partial}}=0}  \nonumber \\
&=&\left. (X^{\underline{m}+(l,v,-2l-v,l)}\,\bar{\triangleright}\,\hat{%
\partial}^{\,\underline{m}})\right| _{\underline{\hat{\partial}}=0}, 
\nonumber
\end{eqnarray}
where we used for the last identity that 
\begin{equation}
\varepsilon (f)=\overline{\varepsilon (f)}=\varepsilon (\bar{f}).
\end{equation}
Applying the formulae 
\begin{eqnarray}
(X^{3})^{n} &=&(X^{3/0}+X^{0})^{n} \\
&=&\sum_{i=0}^{n}\binom{n}{i}(X^{3/0})^{i}(X^{0})^{n-i},  \nonumber \\%
[0.16in]
(\hat{\partial}^{0})^{m} &=&(\hat{\partial}^{3}-\hat{\partial}^{3/0})^{m} \\
&=&\sum_{j=0}^{m}\binom{m}{j}(\hat{\partial}^{3})^{m-j}(-\hat{\partial}%
^{3/0})^{j}  \nonumber
\end{eqnarray}
along with 
\begin{eqnarray}
&&\Big[(X^{+})^{n_{+}}(X^{3/0})^{n_{3/0}}(X^{0})^{n_{0}}(X^{-})^{n_{-}}\,%
\bar{\triangleright}\, \\
&&\left. (\hat{\partial}^{+})^{m_{-}}(\hat{\partial}^{3/0})^{m_{3/0}}(\hat{%
\partial}^{3})^{m_{3}}(\hat{\partial}^{-})^{m_{+}}\Big]\right| _{\underline{%
\hat{\partial}}=0}  \nonumber \\
&=&\Big[(\hat{\partial}^{+})^{n_{+}}(\hat{\partial}^{3/0})^{n_{3/0}}(\hat{%
\partial}^{0})^{n_{0}}(\hat{\partial}^{-})^{n_{-}}\,\bar{\triangleright}\, 
\nonumber \\
&&\left. (X^{+})^{m_{-}}(X^{3/0})^{m_{3/0}}(X^{3})^{m_{3}}(X^{-})^{m_{+}}%
\Big]\right| _{\underline{X}=0}  \nonumber
\end{eqnarray}
to the last expression in (\ref{UmfDulPaar}) gives us the identity 
\begin{eqnarray}
&&\left\langle \hat{\partial}^{\,\underline{m}},X^{\underline{m}%
+(l,v,-2l-v,l)}\right\rangle _{\hat{L},R} \\
&=&\sum_{i=0}^{m_{3}-2l-v}\sum_{j=0}^{m_{3/0}}(-1)^{j}\binom{m_{3}-2l-v}{i}%
\binom{m_{3/0}}{j}  \nonumber \\
&\times& \Big[(\hat{\partial}^{+})^{m_{-}+l}(\hat{\partial}%
^{3/0})^{m_{3/0}+v+i}(\hat{\partial}^{0})^{m_{3}-2l-v-i}(\hat{\partial}%
^{-})^{m_{+}+l}\,\bar{\triangleright}\,  \nonumber \\
&&\left. (X^{+})^{m_{+}}(X^{3/0})^{m_{3}+j}(X^{3})^{m_{3/0}-j}(X^{-})^{m_{-}}%
\Big]\right| _{\underline{X}=0}.  \nonumber
\end{eqnarray}
Since in the case $v>0$ the exponent of $\hat{\partial}^{3/0}$ is strictly
greater than the exponent of $X^{3}$ it follows from (\ref{PropPar3/0}) that 
\begin{equation}
\left\langle \hat{\partial}^{\,\underline{m}},X^{\underline{m}%
+(l,v,-2l-v,l)}\right\rangle _{\hat{L},R}=0,\text{\quad if }v>0.
\end{equation}

Additionally, we get from the above results 
\begin{eqnarray}
\left\langle k,j\right\rangle _{\hat{L},R}^{\underline{m}} &=&\left\langle 
\hat{\partial}^{{}\underline{m}+(l_{k},v_{k},-2l_{k}-v_{k},l_{k})},X^{%
\underline{m}+(l_{j},v_{j},-2l_{j}-v_{j},l_{j})}\right\rangle _{\hat{L},R}
\label{GenProp} \\
&=&\left\langle \hat{\partial}^{{}\underline{m^{\prime }}},X^{\underline{%
m^{\prime }}%
+(l_{j}-l_{k},v_{j}-v_{k},-2(l_{j}-l_{k})-(v_{j}-v_{k}),l_{j}-l_{k})}\right%
\rangle _{\hat{L},R}  \nonumber \\
&=&0,\quad \text{if }v_{j}>v_{k}\text{ or }2l_{j}+v_{j}>2l_{k}+v_{k}, 
\nonumber
\end{eqnarray}
where \underline{$m$}$^{\prime }={}$\underline{$m$}$%
+(l_{k},v_{k},-2l_{k}-v_{k},l_{k}).$ However, the condition in (\ref{GenProp}%
) holds for $k<j$ which immediately gives us 
\begin{equation}
\left\langle k,j\right\rangle _{\hat{L},R}^{\underline{m}}=0,\text{\quad if }%
k<j.
\end{equation}

\item  \textbf{The Proof of (\ref{Propf}):}\newline
By specifying the multi-index \underline{$k$} to 
\begin{equation}
\underline{k}=\underline{m}+(l^{\prime },v^{\prime
},-2l^{\prime }-v^{\prime },l^{\prime })
\end{equation}
with 
\begin{eqnarray}
-\min (m_{+},m_{-}) &\leq &l^{\prime }, \\
-m_{3/0} &\leq &v^{\prime }\leq -1,  \nonumber \\
2l^{\prime }+v^{\prime } &\leq &m_{3},  \nonumber
\end{eqnarray}
the system (\ref{GlSysExpMin}) reduces to 
\begin{eqnarray}
&&\sum_{\QATOP{-m_{3/0}\leq v\leq v^{\prime },-\min (m_{+},m_{-})\leq l}{%
2l+v\leq 2l^{^{\prime }}+{}v^{^{\prime }}}}f_{l,v}^{\underline{m}} \\
&\times& \left\langle \hat{\partial}^{\,\underline{m^{\prime }}},X^{%
\underline{m^{\prime }}+(l-l^{\prime },v-v^{\prime },-2(l-l^{\prime
})-(v-v^{\prime }),l-l^{\prime })}\right\rangle _{\hat{L},R}  \nonumber \\
&=&\delta _{0}^{v^{\prime }}\delta _{0}^{l^{\prime }},  \nonumber
\end{eqnarray}
where \underline{$m$}$^{\prime }={}$\underline{$m$}$+(l^{^{\prime
}},v^{\prime },-2l^{^{\prime }}-v^{^{\prime }},l^{^{\prime }}).$ Notice that
we have a one-to-one correspondence between the equations of the above
subsystem and the pairs $(l^{\prime },v^{\prime })$.
Arranging the equations of this system in an order determined by 
\begin{equation}
(l_{1}^{^{\prime}},v_{1}^{^{\prime }})<(l_{2}^{^{\prime }},v_{2}^{^{\prime
}})\Leftrightarrow \left \{\begin{array}{l} 
 {2l_{1}^{^{\prime }}+v_{1}^{^{\prime}}
<2l_{2}^{^{\prime }}+v_{2}^{^{\prime }},}
\\{2l_{1}^{^{\prime}}+v_{1}^{^{\prime }}
=2l_{2}^{^{\prime }}+v_{2}^{^{\prime }},\quad
v_{1}^{^{\prime }}<v_{2}^{^{\prime }},}\end{array} \right.
\end{equation}
will then show us that it is of triangular form and solves for 
\begin{equation}
f_{l,v}^{\underline{m}}=0,\quad \text{if }2l+v<0\text{ or }v<0.
\end{equation}
\end{enumerate}

\end{document}